\documentclass[titlepage,12pt]{article}
\usepackage{amssymb,psfig,epsfig,pslatex}

\def\shat{\hat{s}}

\def\sp{{\mathbf s}_t}
\newcommand{\sm}{{\mathbf s}_{\bar{t}}}
\newcommand{\kh}{{\hat{\mathbf k}}}
\newcommand{\ph}{\hat{\mathbf p}}
\newcommand{\dhh}{\hat{\mathbf d}}
\newcommand{\one}{1\!\mbox{l}}
\def\percent{\%}
\def\as{{\alpha_{s}}}
\def\Eq#1{Eq.~(\ref{#1})}

\begin{document}
\begin{titlepage}
\noindent
PITHA 01/11 \hfill Nov. 2001 \\
DESY 01-199 \hfill \, \\
TTP01-31    \hfill \, 
\vspace{0.4cm}
\begin{center}
{\LARGE {\bf Top Quark Pair Production and 
Decay including Spin Effects at Hadron Colliders: Predictions at NLO QCD}}\\
\vspace{2cm}
{\bf W. Bernreuther $^{a}$,
A. Brandenburg $^{b,}$,
Z. G. Si $^{a,}$\footnote{Speaker at the conference}
and P. Uwer $^c$}\\
\par\vspace{1cm}
$^a$ Institut f.\ Theoretische Physik, RWTH Aachen, 52056 Aachen, Germany\\
$^b$  DESY-Theorie, 22603 Hamburg, Germany\\
$^c$ Institut f. Theoretische Teilchenphysik, Universit\"at Karlsruhe,
76128 Karlsruhe, Germany
\par\vspace{1.5cm}
{\bf Abstract:}\\
\parbox[t]{\textwidth}
{
Top quark-antiquark ($t\bar t$) pairs will be produced
copiously at the Tevatron collider and in huge numbers at the LHC. 
This will make possible detailed investigations of the properties 
and interactions of this quark flavor. The analysis and interpretation
of future data requires  precise predictions of
the hadronic production of $t\bar t$
pairs and of their subsequent decays.
In this talk the reactions
$p {\bar p}, p p \rightarrow t{\bar t}  + X \rightarrow \ell^+ \ell'^- +
X$ are considered and  results are presented of
our calculation\cite{bbsu:prl}
of the   dilepton angular distribution at
next-to-leading order QCD,  keeping the full
dependence on the spins of the intermediate $t\bar{t}$ state.
The angular distribution is determined for
 different choices of reference axes that can be identified
with the $t$ and $\bar t$ spin axes.
While the QCD corrections to the leading-order distribution
turn out to be small in the case of the LHC, we find them to be
sizeable in the case of  the Tevatron and find, moreover, 
the angular distribution to be
sensitive to the parton content of the proton.
}
\end{center}
\vspace{1cm}
PACS number(s): 12.38.Bx, 13.88.+e, 14.65.Ha

\end{titlepage}

So far the top quark, which was discovered six years
ago\cite{tdisc} is still a relatively
unexplored particle, as compared with other quark flavors.
This will change once the upgraded Tevatron collider
and, in several years, the LHC will be in full operation.
It is expected that about 10$^4$ 
top quark-antiquark ($t\bar t)$
pairs per year will be produced 
at the Tevatron and more than 10$^7$
$t\bar t$ pairs per year at the LHC.
These large data sample
 will make feasible precise investigations of the 
properties and interactions of top quarks.
\par
An important aspect of top quark physics will then come into
play, namely, spin physics with top quarks. It is well-known 
by now that the $t$ quark  is unique among the quark flavors
in that it allows to study
effects associated with its spin in a direct and unambiguous way.
This is due to its  extremely short lifetime 
that prevents the top quark from forming  hadronic
bound states. It behaves like a highly instable ``bare'' quark.
Therefore phenomena
associated with the spins of the top quark and antiquark
are reflected directly
in the distributions and in the corresponding angular correlations 
of the jets, $W$ bosons, or
leptons into which the $t$ and $\bar t$ decay. 
These distributions and correlations reflect the
 $t$ and $\bar t$ polarizations and
spin correlations which in turn
characterize the   $t$ and $\bar t$
production and  decay mechanism(s). 

On the theoretical side the spin
correlations of hadronically produced 
$t\bar t$ pairs were studied  some time 
ago\cite{study,Mahlon:1997,Brandenburg:1996} to leading order in the
coupling $\as$ of Quantum Chromodynamics (QCD).
There exists also an extensive literature,
for example refs.\cite{newphys} and references therein,
on how to exploit top-quark spin phenomena at hadron colliders in the
search for new interactions.
For instance the spin and parity of
a  new heavy resonance that strongly
couples to $t\bar t$ in the s channel could be pinned down with
spin correlations. With appropriate observables that reflect the
$t$ and/or $\bar t$ spins one can, for instance, check whether
or not the $V-A$ law is valid also for top decay, $t\to W b$, or
search for non-standard CP violation in $t\bar t$ production and/or
decay.
A prerequisite of this kind of  
experimental analysis
is that  these spin effects
must  be known as
precisely as possible within the standard model (SM) of particle physics.
Therefore we have determined the production of $t\bar{t}$ pairs by
$q\bar{q}$ annihilation, gluon-gluon fusion and gluon-$q(\bar{q})$ scattering
at next-to-leading order (NLO) in the QCD  
coupling for arbitrary $t$ and $\bar t$
spins\cite{Bernreuther:2000yn}.
Moreover, we have analyzed the  hadronic production
of $t\bar t$ pairs and their subsequent decays  to order $\alpha_s^3$,
keeping the full information on the spins  of the intermediate $t\bar t$ state. 
Within the SM,  where the main top-quark decay
modes are $t\to b W \to b q {\bar q}',  b \ell \nu_{\ell}$, 
the most powerful analyzers of
the polarization of the top quark are the charged leptons, 
or the jets that originate from quarks of weak isospin
$-1/2$ produced by the decay of the $W$ boson. Here we restrict
ourselves to the  channels where  
both $t$ and $\bar t$ decay semileptonically,
\begin{equation}
p {\bar p}, p p \rightarrow {\bar t} t + X \rightarrow \ell^+ \ell'^- + X,
\label{eq:ttsemi}
\end{equation}
$(\ell = e,\mu,\tau)$, 
and we present our predictions of  the
dileptonic angular distribution\cite{bbsu:prl} 
that encodes the $t\bar t$ spin
correlations.
\par
At the parton level the NLO analysis involves  
the following  subprocesses:
\begin{eqnarray}
gg, q{\bar q} & {\buildrel
 t{\bar t}\over \longrightarrow} &  b {\bar b} \ell^+
 \ell'^- \nu_{\ell} {\bar \nu}_{\ell'},
\label{eq:ttrec1}\\
gg, q{\bar q} & {\buildrel
 t{\bar t}\over \longrightarrow} & b {\bar b} \ell^+
 \ell'^- \nu_{\ell} {\bar \nu}_{\ell'} + g,
\label{eq:ttrec2}\\
g + q ({\bar q})&  {\buildrel
 t{\bar t}\over \longrightarrow}&   b {\bar b} \ell^+
 \ell'^- \nu_{\ell} {\bar \nu}_{\ell'} + q ({\bar q}) .
\label{eq:ttrec3}  
\end{eqnarray}
At the Tevatron  $t\bar t$ production
 is dominated by quark-antiquark 
annihilation while at the LHC it is mainly due to gluon-gluon fusion.
\par
Because the total width
$\Gamma_t$
of the top quark is much smaller than its mass $m_t$ 
($\Gamma_t/m_t ={\cal O}(1\%)$), we can 
expand the amplitudes of the parton reactions
(\ref{eq:ttrec1}) - (\ref{eq:ttrec3})
around the poles
of the unstable $t$ and
$\bar t$ quarks and keep only the leading term of this
expansion, i.e.,
the residue of the double poles.  The radiative
corrections to the respective lowest-order amplitudes can be classified
into so-called
factorizable and non-factorizable
corrections.  We take into account  the factorizable
 corrections to the above reactions for which
the squared
matrix element ${\cal M}^{(\lambda)}$ is
of the form
$|{\cal M}^{(\lambda)}|^2 \propto {\rm Tr}[\rho^{(f)}
R^{(a, i)}{\bar\rho}^{(\bar f')}] .$
Here 
$\lambda=1,...,6$ labels the 6 amplitudes
of (\ref{eq:ttrec1}) - (\ref{eq:ttrec3}),
and $R^{(a, i)}$ denotes the respective 
spin density matrix for the production of
on-shell $t\bar t$ pairs.  The superscript 
$a$ labels the initial state and $i$ lables the intermediate state,
i.e., $i=t{\bar t}, t{\bar t}g, t{\bar t}q, t{\bar t}{\bar q}$
state.
The decay density matrix $\rho^{(f)}$(${\bar{\rho}^{(\bar f')}}$) 
describes the normalized angular 
distribution of the decay of a polarized 
$t(\bar t)$ quark into $\ell^+ (\ell^-) + anything$ in the  
rest frame of the $t(\bar t)$ quark. Note that for the reactions
(\ref{eq:ttrec2}) the squared matrix elements
$|{\cal M}^{(\lambda)}|^2$ have, for each $\lambda$,
 three different contributions of
the form ${\rm Tr}[\rho^{(f)} R^{(a, i)}{\bar\rho}^{(\bar f')}]$
because the final-state gluon is either associated with the $t\bar t$
production, the $t$, or $\bar t$ decay amplitude.
\par 
Let us first discuss the production density matrices  $R^{(a, i)}$
and the QCD-induced spin correlation effects at the level of the
$t\bar t$ states. 
The spin density matrix $R^{(a, i)}$ is defined in terms of 
the respective transition matrix element ${\cal T}(a\to i)$.
For instance, for $gg\to t\bar t$ we have 
\begin{equation}
R^{(gg, t\bar t)}_{\alpha\alpha' ,\beta\beta'}=
\frac{1}{N_{gg}}
\sum_{{{\rm\scriptscriptstyle colors} \atop
{\rm\scriptscriptstyle initial}\;
{\rm\scriptscriptstyle spins} }}
\langle t_\alpha\bar t_\beta |{\cal T}|
\,g g\,\rangle\;
\langle \,g g\,|{\cal T}^\dagger|
t_{\alpha'}\bar t_{\beta'}\rangle\;,
\label{eq:Rdef}
\end{equation}
where
the factor $N_{gg}$
averages over the spins and colours
of the initial pair of partons.
The matrix structure of the $R^{(a, i)}$  
is (for ease of notation we drop the superscripts)
\par
\begin{eqnarray}
R_{\alpha\alpha',\beta\beta'}&=&
A \delta_{\alpha\alpha'}\delta_{\beta\beta'}
+B_{i} (\sigma^i)_{\alpha\alpha'}
\delta_{\beta\beta'} +{\bar B}_{i} \delta_{\alpha\alpha'}
(\sigma^i)_{\beta\beta'} \nonumber \\
&&+\, C_{kl}(\sigma^k)_{\alpha\alpha'}
(\sigma^l)_{\beta\beta'}  \,\, ,
\label{eq:Rstruct}
\end{eqnarray}
\par\noindent
The function
$A={\rm Tr}(R)/4$
determines the differential cross section
with $t{\bar t}$ spins summed over.
Because of parity  invariance  the vectors
${\bf B},{\bf\bar B}$ can
have, within QCD,
only a component normal to the scattering plane. This component,
which amounts to a normal polarization of the $t$  and $\bar t$ quarks,
is induced by the absorptive part of the scattering amplitude which,
for the $i=t\bar t$ intermediate state, 
is non-zero  to order $\alpha^3_s$.
The normal polarization is quite small, both for
$t\bar t$ production
at the Tevatron and at the LHC \cite{Bernreuther:1996}.
The functions $C_{kl}$ encode the correlation  between the  $t$ and
${\bar t}$ spins.

In the computation of the  $R^{(a, i)}$ to 
order $\alpha_s^3$   we used
dimensional regularization to treat both the
ultraviolet
and the infrared/collinear singularities.
 Renormalization was performed using
 the $\overline{\rm{MS}}$ prescription for the QCD coupling $\alpha_s$
and the on-shell definition of the top mass $m_t$.
For $a=q{\bar q},gg$ 
the soft and collinear singularities in  $R^{(a, t\bar t)}$, which
appear as single and double poles in $\epsilon = (4-D)/2$,
 are cancelled after including the contributions from  $R^{(a, t{\bar t}g)}$
in the soft and collinear limits
 and  after mass factorization. For the latter we
used the $\overline{\rm{MS}}$ factorization scheme.
The order $\alpha_s^3$  production density matrices of
(\ref{eq:ttrec3})
contain initial state  collinear singularities which are also
removed by mass factorization. The soft and collinear singularities
were extracted by employing
a simplified version of the phase-space slicing technique.
\par
From these density matrices one can obtain, in particular,
the total parton cross sections $\hat\sigma_a(\hat s)$ for the reactions
$g g$,$q\bar{q}$, $q/(\bar{q})g$ $\to t{\bar t} + X$ at NLO.
We have computed these cross sections\cite{Bernreuther:2000yn}
and found   excellent agreement with previous 
results\cite{Nason:1988,Beenakker:1989}.
\par
In order to study the $t\bar t$ spin correlations at the parton level 
we consider the following class of  observables
\begin{equation}
{\cal O}=4\,(\hat{{\bf a}}\cdot {\bf s}_t)  
(\hat{{\bf b}}\cdot {\bf s}_{\bar{t}})
\label{oobse}
\end{equation}
where $\sp,\sm$ denote the $t$ and $\bar t$ spin operators, and 
$\hat{{\bf a}}$ and $\hat{{\bf b}}$ are reference directions that
serve as spin axes. We choose the following vectors:
  \begin{eqnarray}  \label{eq:spbasis}
    \hat {\bf a} = \kh_t,  \hat {\bf b} = \kh_{\bar t} &&
    \mbox{(helicity basis)},\nonumber \\ 
    \hat {\bf a} = \ph,~ \hat{\bf b} = \ph\  &&\mbox{(beam\
      basis)},\nonumber \\
    \hat {\bf a} =\dhh_t, \hat{\bf b} = \dhh_{\bar t} &&
    \mbox{(off-diagonal\ basis)}.
  \end{eqnarray}
Here $\kh_t  (\kh_{\bar t})$ denotes the direction of
flight of the $t (\bar t)$ quark in the parton center-of-mass frame, 
and $\hat{\bf p}$ is 
the unit vector along one of the hadronic beams in the laboratory frame.
Furthermore $\dhh_t$ is the axis  with respect to which the spins
of $t$ and $\bar t$ produced by $q {\bar q}$ annihilation are 100 \percent\
correlated\cite{Mahlon:1997,remark} to leading order in $\alpha_s$.
(For $gg\to t\bar t$ one can show that no spin basis
with this property exists.)

\begin{figure}[htbp!]
\caption{The unnormalized expectation
value of the spin correlation observable (\ref{oobse}) in the beam
basis.
The plots show  
$g^{(0)}_{a}(\eta)$(dotted line), $\ g^{(1)}_{a}(\eta)$(full line) and
${\tilde g}^{(1)}_{a}(\eta)$(dashed line) as a function
of $\eta$ for the  $q\bar{q}$ (a), 
and  $gg$ (b) initial state.}
\centerline{\psfig{figure=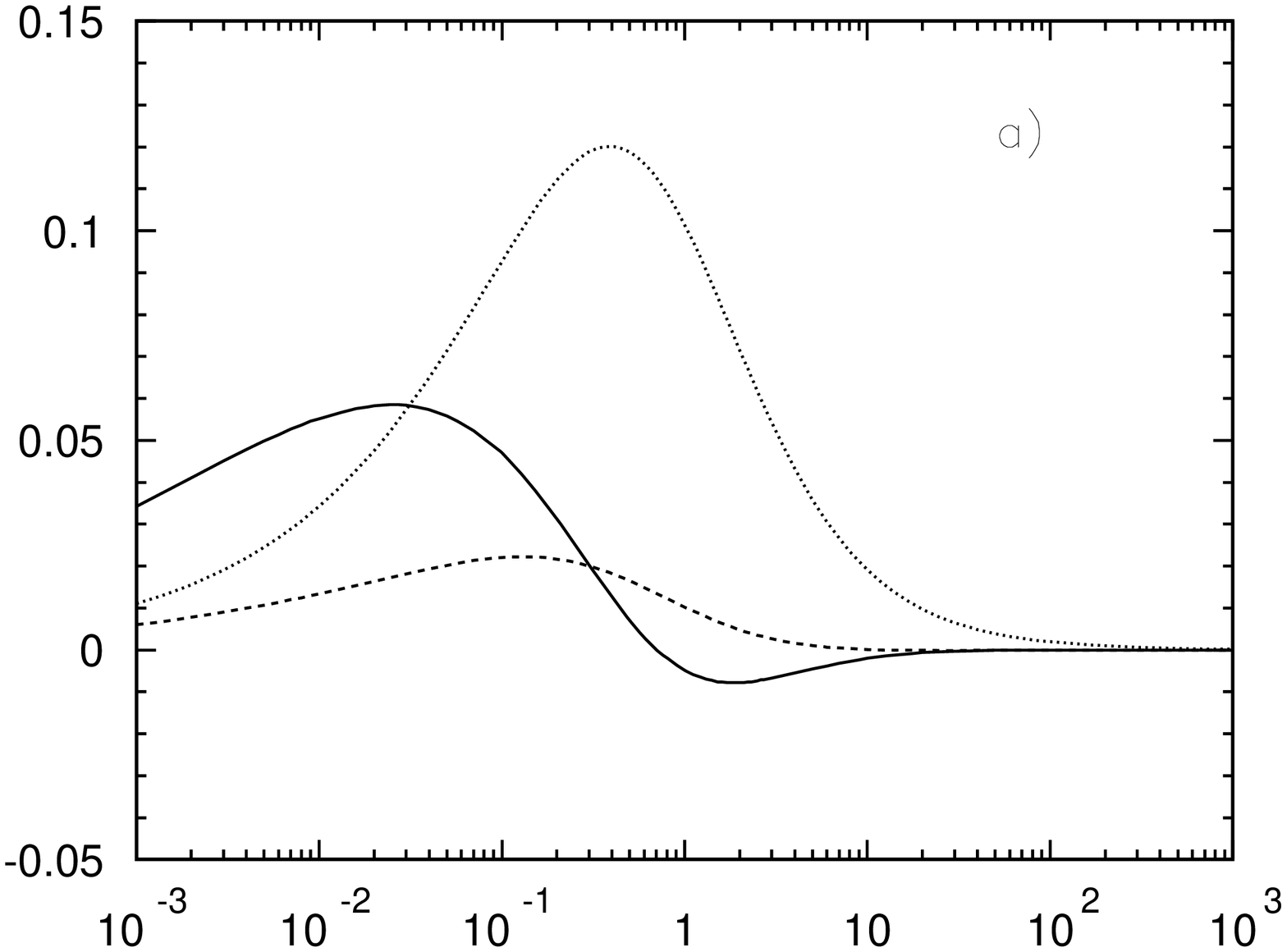,width=5cm}~~~~~~~~~~~
\psfig{figure=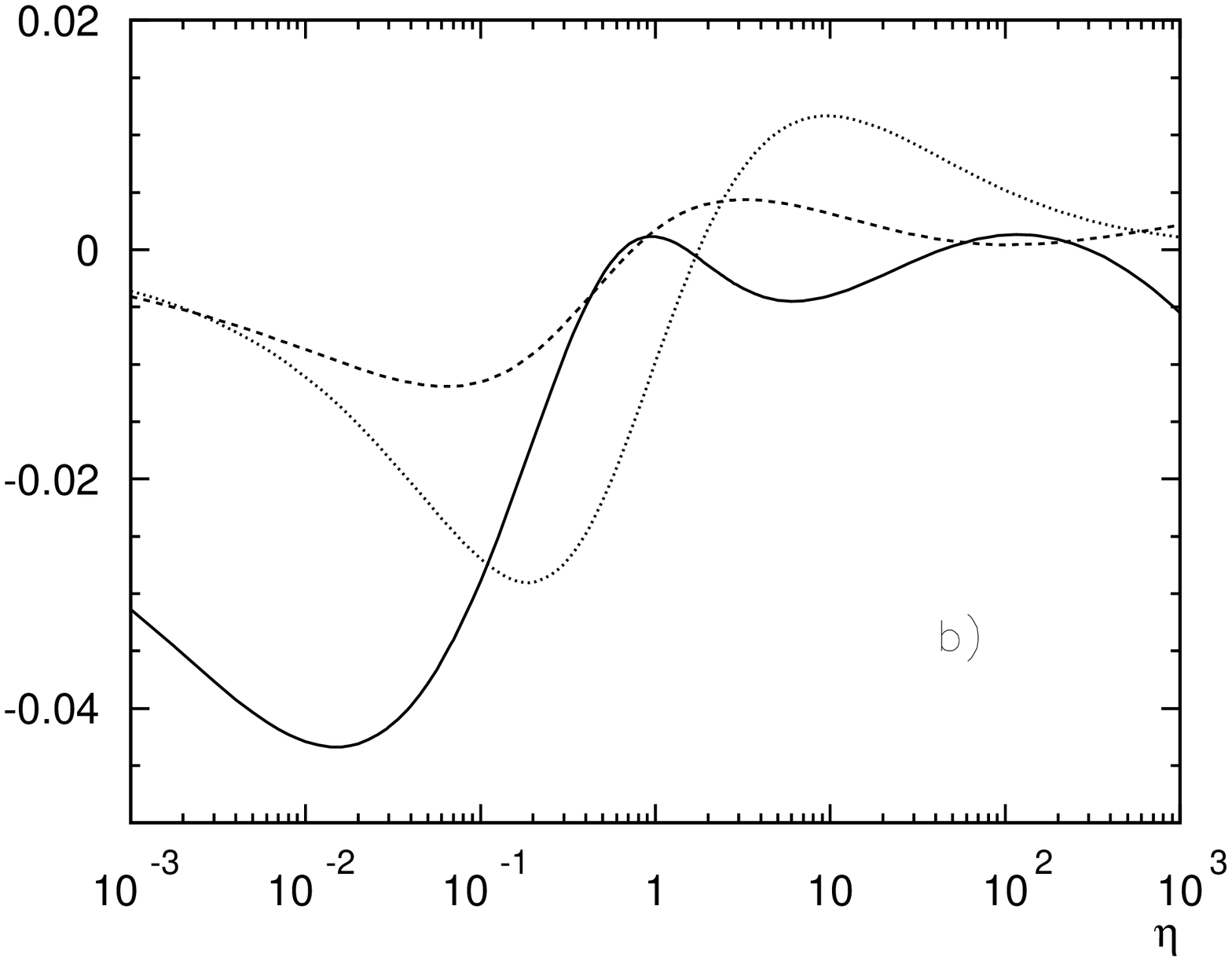,width=5cm}}\label{fig:par}
\end{figure}
The expectation values of the  above observables
${\cal O}$ signify  the degree of correlations among the $t\bar{t}$
spins when using the above  spin quantization axes.
If one identifies the $\overline{\rm{MS}}$ renormalization scale $\mu_R$
with the mass factorization scale $\mu_F$, $\mu_R=\mu_F=\mu$,
  and neglects all quark
masses except for $m_t$, then one can express the  unnormalized 
expectation value of a
spin-correlation observable ${\cal O}$
 in terms of dimensionless scaling functions 
 as follows:
\begin{equation}
{\hat\sigma_a} \langle {\cal O} \rangle_{a}  = \frac{\alpha_s^2}{m_t^2}
[ g^{(0)}_{a}(\eta) + 4\pi\alpha_s(g^{(1)}_{a}(\eta) +
{\tilde g}^{(1)}_{a}(\eta) \ln(\mu^2/m^2_t))] \, ,
\label{eq:expval}
\end{equation}
where $a=g g,q\bar{q}, q/(\bar{q})g$, 
$\shat$ is the parton center-of-mass energy squared, and
$\eta = \frac{\shat}{4m^2_t} -1$.
As an example the functions
$g^{(0)}_{a}(\eta),\ g^{(1)}_{a}(\eta)$ and
${\tilde g}^{(1)}_{a}(\eta)$ are shown in 
Fig.\ref{fig:par} for ($a=q\bar{q},gg$) 
and  the observable in the  beam basis.
The contributions of the $q/(\bar{q})g$ initial states 
to the spin correlations at the hadron level 
are very small. 
The results for the other spin observables are given in 
ref.\cite{Bernreuther:2000yn}.
\par
The QCD corrections to the unnormalized spin correlations are large close
to threshold. This behaviour is due to the factor $\hat{\sigma}_{a}$
which, in this order of perturbation theory, is non-zero at
threshold due to Coulomb attraction.
For hadronic observables these effects are damped by the parton distribution
functions and integration over the momenta  of the partons in the
initial state.
\par
Next we discuss $t$ and $\bar t$ decay. Here only the semileptonic
decays of a polarized $t({\bar t})$ into $\ell^+({\ell'^-})$ +
$anything$ are considered. The decay density matrix 
 $\rho^{(f)}$(${\bar{\rho}^{(\bar f')}}$) 
  has the form
$2\rho^{(f)}_{\alpha'\alpha}
= (\one +{\kappa}_+\,\mathbf{\sigma} \cdot {\hat{\bf{q}}_+})_{\alpha'\alpha}$
where $\hat{\bf{q}}_+$
describes the direction of flight of $\ell^+$ in the rest frame of the $t$
quark and $\sigma_i$ denote the Pauli matrices.  
The decay matrix $\bar{\rho}^{(\bar f')}$ is obtained from $\rho^{(f)}$ 
by replacing  $\hat{\bf{q}}_+$ by $-\hat{\bf{q}}_-$ and
${\kappa}_+$ by ${\kappa}_-$.
The factor ${\kappa}_+$ (${\kappa}_-$) signifies the top-spin 
analyzing power of the charged lepton. It is
equal to one to lowest order in the SM, that is, for $V-A$ charged currents.
Its value including the order $\alpha_s$ corrections can be extracted from
the results of \cite{Czarnecki:1991} and turns out to be very close 
to one: ${\kappa}_+ = {\kappa}_- =1-0.015 \alpha_s$.
\par
So far to the building blocks with which the factorizable
contributions (which are gauge-invariant)
to the squared matrix elements of the above reactions are determined at NLO.
As far as the non-factorizable NLO QCD corrections are concerned which were
calculated in ref.\cite{Beenakker:1999}, we 
expect that their contribution to the spin effects
is considerably smaller than those of the 
factorizable corrections given below.
\par
Now we consider the hadronic  reactions (\ref{eq:ttsemi}) 
and analyze the following double leptonic distribution,
\begin{equation}
\frac{1}{\sigma_t}\frac{d^2\sigma_t}{d\cos\theta_+ d\cos\theta_-}=
\frac{1}{4} (1 -{\rm C}\cos\theta_+ \cos\theta_-)\,\, ,
\label{eq:ddist1}
\end{equation}
with $\sigma_t$ being the cross section for the channel under consideration.
In \Eq{eq:ddist1}  $\theta_+$ ($\theta_-$) denotes the angle between the
direction of flight of the lepton $\ell^+$ ($\ell\,'^-$) in the $t$ ($\bar{t}$)
rest frame and a reference
direction $\hat{\bf a}$ ($\hat{\bf b}$). 
Different choices will yield
different values for the coefficient $C$.
Using the general expressions for
$\rho$, $\bar{\rho}$ and the fact that the factorizable contributions
are of the form ${\rm Tr}[\rho R{\bar{\rho}}]$ we have obtained the
following formula for the correlation coefficient $\rm C$ in 
\Eq{eq:ddist1}:
\begin{equation}
{\rm C} = 4 {\kappa}_+{\kappa}_- \langle (\hat{\bf{a}}\cdot
\sp)(\hat{\bf{b}}\cdot \sm) \rangle   .
\label{eq:ddist2}
\end{equation}
The expectation value
in \Eq{eq:ddist2} is defined with respect to the matrix elements for
the hadronic production of $t\bar t X$.
It can be expressed in terms of  the more familiar double spin asymmetry
\begin{equation}
4 \langle (\hat{\bf{a}}\cdot \sp)(\hat{\bf{b}}\cdot \sm) \rangle = 
\frac{\rm
N(\uparrow \uparrow)+\rm N(\downarrow \downarrow)
  - \rm N(\uparrow \downarrow)- \rm N(\downarrow \uparrow)}{
  \rm N(\uparrow \uparrow)+\rm N(\downarrow \downarrow)
  + \rm N(\uparrow \downarrow)+ \rm N(\downarrow \uparrow)
  },
\label{eq:ddist3}
\end{equation}
where $N(\uparrow \uparrow)$ etc. denote the number
of $t\bar t$ pairs  with $t$ and $\bar t$ spin parallel -- or anti-parallel --
to  $\hat{\bf a}$ and $\hat{\bf b}$, respectively.
From \Eq{eq:ddist3} one can see that 
the axes $\hat{\bf a}$, 
$\hat{\bf b}$ introduced in \Eq{eq:ddist1} through the angles $\theta_\pm$ 
can be interpreted, as in \Eq{oobse}, to be the spin axes 
 of the intermediate $t\bar t$
state within our approximation.
This means that the coefficient C in \Eq{eq:ddist1}
reflects spin correlations of the $t\bar t$ intermediate
state. \Eq{eq:ddist2}  holds for factorizable contributions to all
orders in $\alpha_s$.
\par
\begin{table}[htpb!]
\caption{Coefficient ${\rm C}$
of \Eq{eq:ddist2}
to leading (LO) and
next-to-leading order (NLO)  in $\alpha_s$ for the spin bases of 
\Eq{eq:spbasis}. The CTEQ parton distribution
functions\protect\cite{CTEQ5} were used 
and we have chosen $\mu_R=\mu_F = m_t =$ 175 GeV.}
\par
\begin{center}
{\begin{tabular}{ccccc}\hline
 &\multicolumn{2}{c}{$p\bar p$ at $\sqrt{s}=2$~TeV }
&\multicolumn{2}{c}{$pp$ at $\sqrt{s}=14$~TeV } \\
& LO & NLO & LO & NLO\\  \hline
${\rm C}_{\rm hel.}$ & $-0.456$& $-0.389$ & $\hphantom{-}0.305$  & 
$\hphantom{-}0.311$\\
${\rm C}_{\rm beam}$ & $\hphantom{-}0.910$&  $\hphantom{-}0.806$ & 
$-0.005$ & $-0.072$\\
${\rm C}_{\rm off.}$ & $\hphantom{-}0.918$ & $\hphantom{-}0.813$ & 
$-0.027$ & $-0.089$\\ \hline
\end{tabular}}\label{tab:cteq5}
\end{center}
\end{table}
Table \ref{tab:cteq5}
contains our results for ${\rm C}$ at leading and
next-to-leading order
in $\alpha_s$ using the parton distribution
functions\cite{CTEQ5} CTEQ5L (LO) and CTEQ5M (NLO).
These numbers and the results given
below were obtained by
integrating over the full phase phase.
A further technical comment is in order here. In order to match
with the definition of the QCD coupling used in the evolution of the
PDF, we have expressed the ${\overline{\rm MS}}$ coupling $\alpha_s$
of 6 flavor QCD by the ${\overline{\rm MS}}$ coupling whose evolution
is governed by the beta function that depends only on the 5 light quark
flavors.
\par
For $p \bar p$ collisions at $\sqrt{s}=$ 2 TeV
the helicity basis is not
the best choice  because the $t$, $\bar t$
quarks are only moderately relativistic in this case.
Table  \ref{tab:cteq5} shows that the dilepton spin correlations
at the Tevatron are large both in  the
off-diagonal and in the beam basis. These two spin bases 
yield almost identical results.
The QCD corrections
decrease the LO  results for these correlations
by about 10\percent. Since the $gg$ initial state
dominates $t\bar t$ production
with  $p p$ collisions at 
$\sqrt{s}=$ 14 TeV  the beam and off-diagonal bases
are no longer useful.
Here the helicity
basis is a good choice and gives a spin correlation of about 30\percent.
In this case the QCD corrections are small.
The large difference between the LO and NLO results for the correlation
in the beam basis
at the LHC is due to an almost complete cancellation of the
contributions from the $q\bar{q}$ and $gg$ initial state at LO. 
\par
As usual, there are several sources of theoretical errors
at fixed order in perturbation theory.
As to the scale uncertainty,  the inclusion of the QCD corrections
reduces the dependence of the $t\bar{t}$ cross section $\sigma_t$ on
the renormalization and factorization scales significantly.
The same is true
for  the product $\sigma_t {\rm C}$ -- see ref.\cite{bbsu:prl}
for details.
\begin{table}[htb]
\caption{Upper part: Dependence of the correlation coefficients,
computed with the PDF  of ref.\protect\cite{CTEQ5}, and
$\mu=\mu_R=\mu_F$ at NLO. Lower part:
Correlation coefficients ${\rm C}_{\rm hel.}$,
${\rm C}_{\rm beam}$, and ${\rm C}_{\rm off.}$
at NLO for $\mu_R=\mu_F=m_t$ and different sets
of parton distribution functions:  GRV98\protect\cite{Gluck:1998xa}, 
CTEQ5\protect\cite{CTEQ5}, and MRST98 (c-g)\protect\cite{MRST98}.} 
\par
\begin{center}
{\begin{tabular}{ccccc} \hline
& \multicolumn{3}{c}{$p\bar p$ at $\sqrt{s}=2$~TeV}
& $p p$ at $\sqrt{s}=14$~TeV \\
 $\mu_R=\mu_F$   & ${\rm C}_{\rm hel.}$ &  ${\rm C}_{\rm beam}$ &
 ${\rm C}_{\rm off.}$ &  ${\rm C}_{\rm hel.}$ \\ \hline
$m_t/2$ & $-0.364$   & 0.774    & 0.779  &   0.278
\\
$m_t$   & $-0.389$   & 0.806  & 0.813 &  0.311
\\
$2m_t$ & $-0.407$  & 0.829 & 0.836  &  0.331
\\ \hline \hline
 PDF   & ${\rm C}_{\rm hel.}$ &  ${\rm C}_{\rm beam}$ &
 ${\rm C}_{\rm off.}$ &  ${\rm C}_{\rm hel.}$ \\ \hline
GRV98 & $-0.325$   & 0.734    & 0.739  &  0.332
\\
CTEQ5   & $-0.389$   & 0.806  & 0.813 &  0.311
\\
MRST98 & $-0.417$  & 0.838 & 0.846  &  0.315
\\ \hline
\end{tabular}}\label{tab:mudep}
\end{center}
\end{table} 
To leading order in $\alpha_s$ the coefficient C depends only
on the factorization scale $\mu_F$, while at NLO  it
depends on both scales $\mu_R$ and $\mu_F$. Table \ref{tab:mudep}
shows our NLO results for the three choices $\mu_R=\mu_F=m_t/2,m_t,2m_t$,
again using the PDF of ref.\cite{CTEQ5}.
\par
In Table \ref{tab:mudep}
we also compare our results for C obtained with
different sets of PDF.
For  $p\bar{p}$ collisions at $\sqrt{s}=2$ TeV,
the spread of the results is larger
than the scale uncertainty given in the upper part of
Table \ref{tab:mudep}.
To a considerable extent this is due to an important property  of C,
namely the $q\bar{q}$ and $gg$ initial states
contribute to C with opposite signs.
Therefore the spin correlations are quite sensitive
to the relative weights of $q\bar{q}$ and $gg$ initiated $t\bar t$ events.
These weights depend in particular on the chosen set of PDF.
For example, we find the following individual NLO contributions
for the helicity, beam, and off-diagonal correlation 
at the upgraded Tevatron: for the GRV98 (MRST98) PDF 
C$^{q\bar q}_{\rm hel.}=-0.443\ (-0.486)$, 
C$^{gg}_{\rm hel.}=+0.124\  (+0.075)$, 
C$^{q\bar q}_{\rm beam}=+0.802\ (+0.879)$, 
C$^{gg}_{\rm beam}=-0.068\  (-0.042)$, and
C$^{q\bar q}_{\rm off.}=+0.810\ (+0.889)$, 
C$^{gg}_{\rm off.}=-0.073\  (-0.044)$. This suggests that
accurate measurements  of the dilepton
distribution (\ref{eq:ddist1}), using different spin bases,
at the upgraded Tevatron could be a useful tool in the effort
to improve the knowledge of the PDF.
\par
The above analysis can be extended
 to the ``lepton+jets'' and ``all jets'' decay
channels in a  straightforward fashion\cite{bbsu3}. 
The ``lepton+jets'' channels should be particularly useful
for detecting $t\bar t$ spin correlations: although one looses top-spin
analyzing power one gains in statistics and the experimental 
reconstruction of the $t$ and $\bar t$ rest frames may also be facilitated.
\par
To conclude: we  have determined, at NLO in the QCD coupling, 
the dileptonic angular distribution (\ref{eq:ddist1})  that reflects
the degree of correlation between the $t$ and $\bar t$  spins.
Our  results for the Tevatron
show that the scale and in particular the
PDF uncertainties in the prediction
of this
distribution  must be reduced before $t\bar t$ spin
correlations can be
used in a meaningful way to search for
relatively small effects of new interactions affecting
$t\bar t$ production and/or decay, but respect, like QCD, parity and CP.
 For example, the effect of a small anomalous 
chromomagnetic $t{\bar t}g$ coupling that would lead to
deviation from the SM predictions at the, say,  few percent level 
would be swamped by these uncertainties.  
On the other hand our Tevatron results should  be useful to
learn more about the parton distributions in the proton at high energies.
For $pp$ collisions at $\sqrt{s}$ = 14 TeV the theoretical uncertainties
in the prediction of this distribution are, fortunately, 
smaller. This gives rise to expectations that 
top quark spin correlations will play an important role in the 
precision  analysis of $t\bar t$ events at the  LHC.     

\section*{Acknowledgements}
This work is supported by BMBF, contract 05 HT1 PAA 4.
A.B. is supported by a Heisenberg grant of the D.F.G.

\end{document}